\documentclass{article}
\usepackage{spconfa4,amsmath,graphicx}

\usepackage[ruled, boxed]{algorithm2e}
\usepackage{xcolor}
\usepackage{multirow, bm, amssymb}
\usepackage{booktabs}
\usepackage{comment}

\title{A lightweight hybrid framework for acoustic echo cancellation and suppression via constrained recursive training}
\name{Huawei Zhang*, Rilin Chen†, Hao Zhang, Meng Yu, Dong Yu\thanks{*This work was done during an internship at Tencent AI Lab.\\†Rilin Chen is the corresponding author: rilinchen@tencent.com.}}
\address{AI Lab, Tencent}
\begin{document}
\ninept
\maketitle
\begin{abstract}
We propose a hybrid framework that combines deep learning with adaptive filtering for acoustic echo cancellation (AEC) and suppression.
It integrates a multi-output neural network to jointly update filter parameters and perform post-processing. 
A learnable variant of the normalized least mean squares (NLMS) filter is introduced, incorporating a variable step size and a variable transition factor that are recursively updated for dynamic adaptation. 
In addition, an enhanced near-end speech estimate is generated from the filter output to suppress the residual echoes. 
To ensure training stability, we propose a constrained recursive training strategy that aligns with the adaptive nature of the filter, where an upper-bound constraint is introduced on the filter output. 
For efficiency, the framework adopts a lightweight implementation with only $253$ k parameters. 
Experimental results show that the framework outperforms the baselines in terms of both signal quality and speech intelligibility.
\end{abstract}

\section{Introduction}
\label{sec:intro}
Acoustic echo cancellation (AEC) \cite{duttweiler2000proportionate, enzner2006frequency, zhang2022neural, ristea2023deepvqe, jiang2024two, wang2021weighted, zhang2022deep, zhang2023neuralkalman, jiang2025small, 11462524, zhang2024enhanced} is a fundamental component in modern speech communication systems, aiming to remove echoes caused by acoustic coupling between loudspeakers and microphones.
Without effective AEC, the echo may severely disrupt the natural flow of conversations.

Adaptive filters \cite{farhang2013adaptive} have long been employed for acoustic echo cancellation, with the normalized least mean square (NLMS) filter \cite{duttweiler2000proportionate} and the Kalman Filter \cite{enzner2006frequency} being the most common choices.
The NLMS filter is computationally efficient and easy to implement, but its performance degrades in practical scenarios.
The Kalman filter, on the other hand, provides robustness to double-talk and faster convergence, yet its effectiveness depends on accurate covariance estimation, which is often difficult to achieve in real-world conditions.
More recently, deep learning–based approaches \cite{zhang2022neural, ristea2023deepvqe, jiang2024two} have been introduced for AEC and suppression.
When trained on sufficiently diverse data, these models can outperform traditional filters, especially in handling nonlinear echo.
However, their ability to generalize to unseen acoustic conditions remains uncertain.

To overcome these limitations, hybrid methods that combine deep learning with adaptive filtering have attracted increasing attention \cite{wang2021weighted, zhang2022deep, zhang2023neuralkalman, jiang2025small, zhang2024enhanced}.
Such methods aim to exploit the complementary strengths of both approaches to achieve superior performance and robustness. 
One class of hybrid methods \cite{zhang2022deep, zhang2023neuralkalman, zhang2024enhanced} enhances adaptive filters by using neural networks to model the non-linearity and estimate key parameters, such as the step size in the NLMS filter or the transition factor in the Kalman filter. 
Another class \cite{wang2021weighted, jiang2025small} employs neural networks for post-filtering, suppressing residual echoes in the filter output.
Despite these advances, existing hybrid methods still face challenges in balancing adaptability, stability, and efficiency, leaving room for further improvement.

Although existing hybrid methods have demonstrated promising performance, their potential has not been fully explored in several key aspects:
(1) Since overall performance strongly depends on the underlying filter design, there remains a need to develop more advanced adaptive filters for hybrid methods.
(2) Simply combining the two types of hybrid methods described above may seem straightforward, but applying two separately trained networks does not guarantee optimal results.
(3) Most hybrid methods combine the adaptive filter with the offline-trained neural networks \cite{zhang2022deep, zhang2023neuralkalman}, which introduces a mismatch between training and the inherently adaptive nature of the filter, thereby limiting effectiveness.
(4) Many methods also require a relatively large number of network parameters, largely due to the need for complex nonlinear modeling to estimate or suppress nonlinear echo.

In this paper, we propose a hybrid framework that further advances AEC and suppression by fully exploiting the synergy between deep learning and adaptive filtering.
The main contributions are summarized as follows:
(1) To integrate the strengths of both NLMS and Kalman filters, we design a learnable variant of the NLMS filter, termed VS-TF NLMS filter, with a variable step size and a variable transition factor, enabling dynamic adaptation.
(2) A single neural network is trained to jointly update filter parameters and perform post-processing, ensuring optimal performance.
(3) To mitigate the mismatch between offline training and online adaptation, we propose a constrained recursive training strategy, in which the filter output is bounded and recursively fed into the network.
(4) For efficiency, we present a lightweight implementation with only a limited number of parameters, realized through a grouped temporal convolutional recurrent network (GTCRN) \cite{rong2024gtcrn}.

\section{Acoustic echo cancellation and suppression}
\label{sec:AEC}

\subsection{Problem Formulation}
\label{ssec:problem_formulation}
In a typical acoustic echo scenario, the time-frequency domain microphone signal $Y(m, k)$ is represented by
\begin{equation}
\label{equ:y_time_frequency}
    Y(m, k) = D(m, k) + S(m, k) + N(m, k),
\end{equation}
where $D(m, k)$, $S(m, k)$, and $N(m, k)$ denote the acoustic echo, the near-end speech, the background noise, respectively.
The frequency bin index $k$ is omitted for brevity in the rest of this paper. 

With the far-end signal $X(m)$ as the reference signal, the goal of this work is to extract the near-end speech $S(m)$ from the received microphone signal $Y(m)$.

\subsection{Related work}
\label{ssec:related}
As shown in \cite{duttweiler2000proportionate}, the classical NLMS filter \cite{farhang2013adaptive} can be implemented to extract the near-end speech $S(m)$ from the received microphone signal $Y(m)$.
It requires a $L$-order filter $\mathbf{W}(m) = [W(m), W(m-1), \cdots, W(m-L+1)]^T$, which can be updated by
\begin{equation}
\label{equ:w_nlms}
\mathbf{W}(m+1) = \mathbf{W}(m) + \mu \frac{\mathbf{X}(m) E(m)}{\mathbf{X}^{H}(m) \mathbf{X}(m)},
\end{equation}
where $\mu$ is the step size, $\mathbf{X}(m)$ is the buffer of the reference signal $X(m)$, and $E(m)$ is the error signal as an estimation of the near-end speech $S(m)$.
Such NLMS filter-based AEC solution can be conveniently implemented with a low calculation cost, but its performance is generally limited in severe scenarios.

By contrast, the Kalman filter-based AEC algorithm \cite{enzner2006frequency} often achieve a better performance, where the $L$-order filter $\mathbf{W}(m)$ is updated by \cite{farhang2013adaptive}  
\begin{equation}
\label{equ:w_kf}
\mathbf{W}(m+1) = A \left[ \mathbf{W}(m) + \mathbf{K}(m) E(m) \right],
\end{equation}
where $A$ is the transition factor and $\mathbf{K}(m)$ is the Kalman gain.
However, its performance replies on the accuracy of estimated covariance matrices.

In \cite{yang2023low, zhang2024neural}, the estimation of covariance matrices for the Kalman filter is achieved by deep learning, but a large calculation cost will be introduced. 

In \cite{zhang2023neuralkalman}, a learnable variant of the Kalman filter was proposed, compensating the inaccurate estimation of covariance matrices by modeling a nonlinear transition function using deep learning.
Experimental results demonstrate that compared with learning-based covariance matrix estimation, such variant achieves a comparable AEC performance with a less calculation cost.
However, there is the lack of theoretical support and interpretability in the design of such variant.

In \cite{zhang2024neural, zhang2024advancing}, a recursive training strategy was proposed for acoustic howling suppression (AHS), aiming to satisfy the recursive nature of acoustic howling.
However, as discussed in \cite{zhang2024neural, zhang2024advancing}, such recursive training may cause the training instability.
Recently, the recursive training strategy was introduced for AEC \cite{11462524}, but it focuses on estimating the filter parameter with the cost of the inability to deal with the non-linear echo to some extent.

\section{The proposed hybrid framework}
\label{sec:framework}

\begin{figure}[t]
\includegraphics[width=8.5cm]{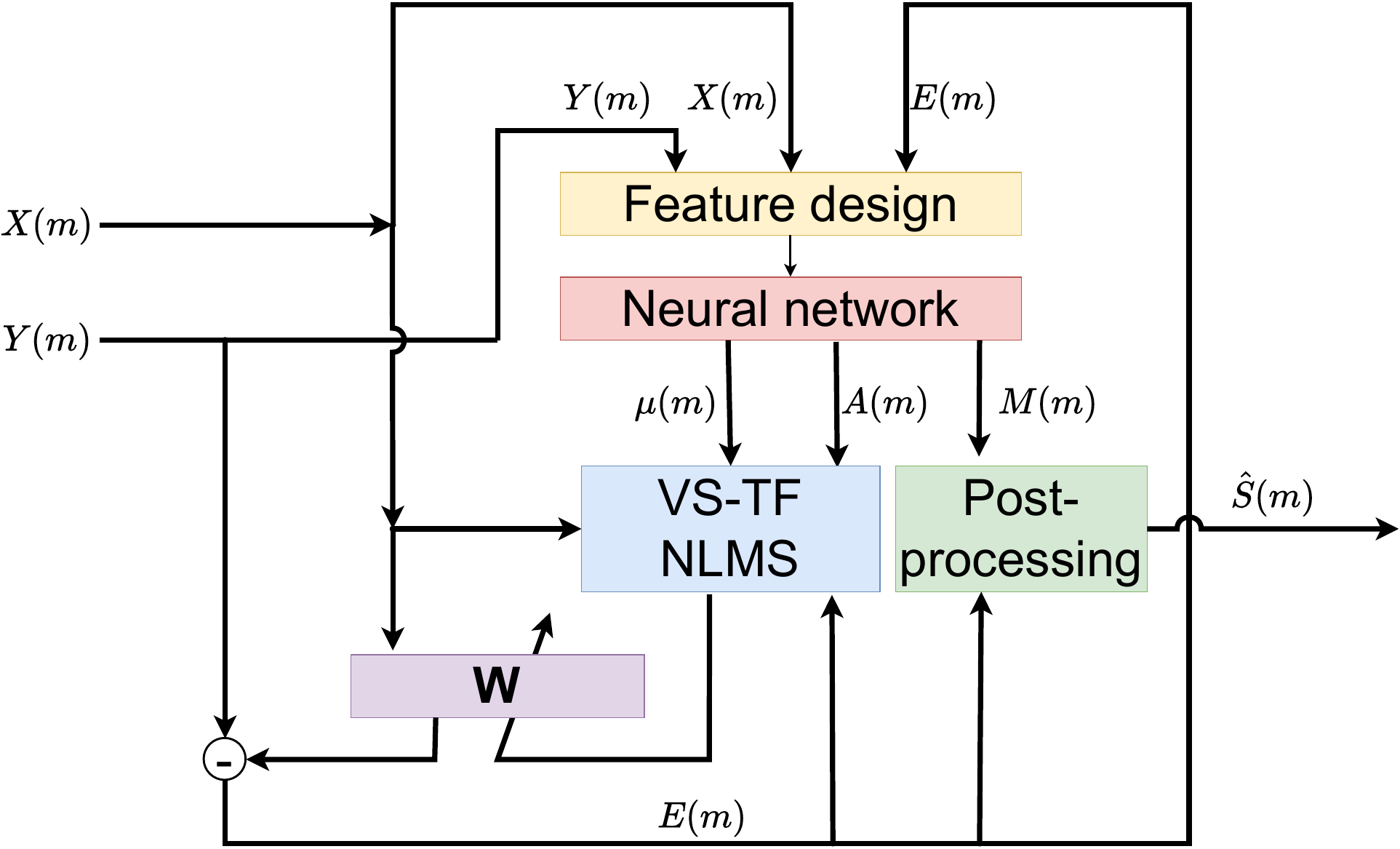}
\centering
\caption{The block diagram of the proposed hybrid framework for acoustic echo cancellation and suppression.}
\label{fig:nn_block}
\end{figure}

In this section, as shown in Fig.~\ref{fig:nn_block}, we propose a hybrid framework to achieve the AEC and suppression, by applying a single neural network to jointly update filter parameters and perform post-processing.

In detail, a lightweight neural network is implemented, consists of a linear layer with Rectified Linear Unit (ReLU) activation, a norm layer, a GTCRN \cite{rong2024gtcrn}, and three separate linear layers with sigmoid activation.

The input feature is a concatenation of three components: $\log(|Y(m)|^2)$, $\log(|X(m)|^2)$, and $\log(|E(m)|^2)$, which are calculated from the microphone signal $Y(m)$, the reference signal $X(m)$, and the error signal $E(m)$ from the adaptive filter, respectively.
The inputs to the neural network are the input feature above and the previous state $h(m-1)$ of the RNN.
The outputs of the neural network include a step size ${\mu}(m)$ and a transition factor ${A}(m)$ for the adaptive filter, and a ratio mask $M(m)$ for the post-processing, and the current state $h(m)$ of the RNN.

\subsection{Variable step-size and transition-factor (VS-TF) NLMS}
\label{ssec:filter}
First, we propose the VS-TF NLMS filter with a variable step size ${\mu}(m)$ and a variable transition factor ${A}(m)$.
With the reference signal $X(m)$, an error signal $E(m)$ is obtained as an initial estimation of the near-end speech $S(m)$.

Based on the theoretical analysis in \cite{enzner2006frequency, yang2017frequency}, we can represent the Kalman filer as the VS-TF NLMS filter with a variable step size $\mu(m)$ and a variable transition factor $A(m)$, where its update rule as Equ.~(\ref{equ:w_kf}) can be rewritten as
\begin{equation}
\label{equ:w_nlms_new}
\mathbf{W}(m+1) = A(m) \left[ \mathbf{W}(m) + \mu(m)  \frac{\mathbf{X}(m) E(m)}{\mathbf{X}^{H}(m) \mathbf{X}(m)} \right],
\end{equation}
where $\mathbf{X}(m)$ is the buffer of the reference signal $X(m)$.
Although the estimation of covariance matrices is removed, there is still the lack of a convenient way to estimate $\mu(m)$ and $A(m)$ in Equ.~(\ref{equ:w_nlms_new}).

To further solve this issue, we introduce deep learning to estimate the variable step size $\mu(m)$ and the variable transition factor $A(m)$ for each iteration, resulting in the VF-TF NLMS filter. 
With the estimation of $\mu(m)$ and $A(m)$ from the neural network, the filter $\mathbf{W}(m)$ can be updated as Equ.~(\ref{equ:w_nlms_new}), achieving a satisfied error signal $E(m)$ by 
\begin{equation}
\label{equ:e}
E({m)} = Y(m) - \mathbf{W}^H(m) \mathbf{X}(m).
\end{equation}

\subsection{Post-processing}
\label{ssec:post_process}
Second, we utilize the neural network to perform post-processing, by suppressing the residual echo in the error signal $E(m)$.
As a result, the output of the whole hybrid framework is obtained as $\hat{S}(m)$, which is the final estimation of the near-end speech $S(m)$.

In detail, a complex ratio mask $M(m)$ \cite{williamson2015complex, wang2018supervised} is estimated, achieving $\hat{S}(m)$ based on the error signal $E(m)$ by
\begin{equation}
\label{equ:s_hat}
\hat{S}(m) = M(m) E(m).
\end{equation}

More details of the proposed framework can be found in Algorithm 1, where $\mathcal{M}$ is the total number of frames, and $\mathbb{NN}[\quad]$ represents the neural network used.

\begin{algorithm}
\caption{The proposed hybrid framework for AEC and suppression}
\KwIn{$Y(m), X(m)$}
\KwOut{$\hat{S}(m)$}
\While{$m \leq \mathcal{M}$} {
$\mathbf{X}(m) \leftarrow {X}(m)$ \;
$E(m) = Y(m) - \mathbf{W}^H(m) \mathbf{X}(m)$ \;
${\mu}(m), {A}(m), M(m) \leftarrow \mathbb{NN}[Y(m), X(m), E(m)]$ \;
$\mathbf{W}(m+1) = {A}(m) \left[ \mathbf{W}(m) + \frac{ {\mu}(m)  \mathbf{X}(m) E(m)}{\mathbf{X}^{H}(m) \mathbf{X}(m)} \right]$ \;
$\hat{S}(m) = M(m) E(m)$ \;
}
\end{algorithm}

\section{Training strategy}
\label{sec:training_strategy}

In this section, we discuss the training strategy for the proposed hybrid framework with the loss function design.

\subsection{Recursive training}
\label{ssec:training_recursive}

We introduce a recursive training strategy  \cite{11462524, zhang2024neural, zhang2024advancing} that aligns with the adaptive nature of the filter.

The offline-trained model in \cite{zhang2022deep, zhang2023neuralkalman} only fed the microphone signal $Y(M)$ and the reference signal $X(m)$ into the neural network, without considering the adaptive nature of the filter.
To avoid the mismatch, we further fed the error signal $E(m)$ into the network as shown in Algorithm 1, providing information related to the adaptive process.
Correspondingly, a recursive training strategy is developed, which applies the previous network outputs to update the current filter $\mathbf{W}(m)$, and then feds the current filter output $E(m)$ into the network recursively during the training.

\subsection{Upper-bound constraint}
\label{ssec:upper_bound_constraint}

However, the training instability may be introduced by such recursive training strategy in Sec.~\ref{ssec:training_recursive}, as discussed in \cite{zhang2024neural, zhang2024advancing}.
To solve this issue, we introduce an upper-bound constraint on the filter output $E(m)$.
As the near-end speech $S(m)$ is one part of the microphone signal $Y(m)$, the filter output $E(m)$ as an estimation of $S(m)$ should not have larger amplitude than $Y(m)$.
Therefore, the filter output $E(m)$ is constrained by the microphone signal $Y(m)$ as
\begin{equation}
\label{equ:e_constrained}
E(m) = 
\begin{cases} 
Y(m), & |E(m)| > |Y(m)|, \\
E(m), & |E(m)| \leq |Y(m)|.
\end{cases}
\end{equation}
With this constraint, we can avoid the training instability by limiting the value range of the input signals to the neural network during the training. 
Correspondingly, the constraint is also introduced during the streaming inference after training.

\subsection{Loss function}
\label{ssec:loss_function}
To enhance the robustness of the proposed hybrid framework, the outputs of the filter and the post-process are constrained by two sub loss functions, respectively.
Both are defined to as
\begin{equation}
\label{equ:loss_1}
L_1 = -\text{SI-SNR}(s, e) + \alpha_1 \text{EAL}(S, E) + \alpha_2 \text{MAE}(S, E),
\end{equation}
\begin{equation}
\label{equ:loss_2}
L_2 = -\text{SI-SNR}(s, \hat{s}) + \alpha_1 \text{EAL}(S, \hat{S}) + \alpha_2 \text{MAE}(S, \hat{S}),
\end{equation}
where SI-SNR~${(\quad)}$ denotes Scale-Invariant Signal-to-Noise Ratio (SISNR) \cite{luo2018tasnet}, EAL~${(\quad)}$ denotes Echo-Aware Loss \cite{zhang2022multi}, and MAE~${(\quad)}$ denotes Mean Absolute Error of complex and magnitude spectrogram between the target and the outputted signals, $e$ and $\hat{s}$ are the time-domain representations of the error signal and the final estimation of the near-end speech, respectively, and $\alpha_1 = 0.04$ and $\alpha_2 = 0.005$ are two parameters to balance three items in each sub loss.
Here, we omit $t$ and $m$ for brevity.

The overall loss is used for the backward propagation during the training, defined as
\begin{equation}
\label{equ:loss_all}
L = \beta*L_1 + (1-\beta) L_2,
\end{equation}
where $\beta = 0.25$ is a trade-off parameter.

\begin{table}[!t]
\centering
\caption{Performance comparison and model size of three proposed frameworks. 
Bold indicates the best score.}
\label{performance_comparison_self}
\begin{tabular}{c c c c c}
\toprule
Method & SI-SNR & SDR & PESQ & \#Params \\
\midrule
Initial state & $1.00$ & $2.47$ & $2.31$ & -- \\
\midrule
NLMS & $1.35$ & $3.14$ & $2.34$ & -- \\
KF & $2.30$ & $3.86$ & $2.56$ & -- \\
\midrule
VS-TF NLMS & $2.57$ & $4.08$ & $2.62$ & \bm{$67$}~\textbf{\text{k}} \\
Half-recursive & \bm{$14.18$} & \bm{$14.38$} & $2.90$ & $168$~{\text{{k}}} \\
Recursive & $13.92$ & $14.14$ & \bm{$3.00$} & $253$~{\text{{k}}} \\
\bottomrule
\end{tabular}
\end{table}

\begin{table}[!t]
\centering
\caption{Performance comparison and model size of the proposed recursive framework and existing hybrid methods. 
Bold indicates the best score.}
\label{performance_comparison}
\begin{tabular}{l c c c}
\toprule
Method & ERLE & PESQ & \#Params \\
\midrule
Initial state & -- & $2.02$ & -- \\
\midrule
NLMSNet & $7.28$ & $2.20$ & $888$~\text{k} \\
KalmanNet & $9.89$ & $2.22$ & $977$~\text{k} \\
\midrule
Recursive & \bm{$25.52$} & \bm{$2.68$} & \bm{$253$}~\textbf{\text{k}} \\
\bottomrule
\end{tabular}
\end{table}

\section{Experimental study}
\label{sec:experiment}

\subsection{Data preparation}
\label{sec:data}
Three simulated datasets are used, including a training set, a validation set and an evaluation set.   
Each utterance lasts $8$~s second, recording the mix of the near-end speech, the acoustic echo, and background noises.
The training set and the validation set are simulated as follows:
The near-end speech is obtained from a Chinese corpus, AISHELL-2 \cite{du2018aishell}, and an English corpus, Librispeech \cite{panayotov2015librispeech}.
The environmental noises are obtained from the DNS-Challenge \cite{reddy2021interspeech}.
We simulate the echo data by convolving the clean speech with room impulse responses (RIRs) from the SLR28 dataset \cite{ko2017study}.
The evaluation set includes $100$ utterances from the AEC challenge dataset \cite{sridhar2021icassp}, whose signal-to-echo ratio (SER) is varying from $-10$ dB to $10$ dB.
The sampling frequency is $16000$~Hz.

\subsection{Experimental setting}
To evaluate the performance of our framework, we compare these methods as the followings:

(1) NLMS: The classical NLMS filter-based method \cite{duttweiler2000proportionate}.

(2) KF: The classical Kalman filter-based method \cite{enzner2006frequency}.

(3) NLMSNet: A hybrid method that integrates the NLMS filter and a large-scale LSTM network \cite{zhang2022deep}.

(4) KalmanNet: A hybrid method that integrates the Kalman filter and a large-scale LSTM network \cite{zhang2023neuralkalman}.

(5) VS-TF NLMS: The proposed VS-TF NLMS filter-based method.
A single-layer recurrent neural network (RNN) is required to estimate the step size $\mu(m)$ and the transition factor $A(m)$.
The training strategy is the proposed constraint recursive training in Sec.~\ref{ssec:training_recursive} and Sec.~\ref{ssec:upper_bound_constraint}, where the loss function is as same as Equ.~(\ref{equ:loss_1}).

(6) Half-recursive:
As a variant of our proposed framework, we use the recursively trained single-layer RNN in (5) to aid the proposed VS-TF NLMS filter first, and then a GTCRN is used to suppress residual echoes.
With the filter output as the input, the training strategy for the GTCRN is offline training like \cite{wang2021weighted}, where the loss function is as same as Equ.~(\ref{equ:loss_2}).

(7) Recursive: The proposed hybrid framework with a GTCRN, as discussed in Sec.~\ref{sec:framework} and Sec.~\ref{sec:training_strategy}.

As a fair comparison, we train neural networks in the proposed methods and the baselines based on the same loss function as Equ.~(\ref{equ:loss_all}).
The Short-Time Fourier Transform (STFT) is implemented with the frame size as $320$ and $50\%$ overlap. 

We evaluate performance in terms of four metrics, including SI-SNR \cite{luo2018tasnet}, Source-to-Distortion Ratio (SDR) \cite{le2019sdr}, Echo Return Loss Enhancement (ERLE) \cite{theodoridis2013academic}, and Perceptual Evaluation of Speech Quality (PESQ) \cite{rix2001perceptual}.

\subsection{Experimental results}
\label{sec:results}

First, three proposed frameworks are built for the ablation study, in terms of SI-SNR, SDR, and PESQ, based on the validation set, as summarized in Table~\ref{performance_comparison_self}.
Compared with the classical NLMS method, the VS-TF NLMS filter improves SI-SNR by approximately $1.2$~dB and SDR by about $0.9$~dB, and also outperforms the Kalman filter (KF) method.
In terms of speech intelligibility, the VS-TF NLMS filter achieves the best performance among the conventional adaptive filter-based methods, yielding a higher PESQ score.
Building upon the VS-TF NLMS filter, the proposed recursive frameworks further improve the performance.
Specifically, the recursive framework provides an additional improvement of about $11.5$~dB in SI-SNR and roughly $10$~dB in SDR compared with the VS-TF NLMS filter, achieving performance comparable to that of the half-recursive framework.
In terms of speech intelligibility, the recursive framework increases PESQ by nearly $0.4$ over the VS-TF NLMS filter, while the half-recursive framework provides smaller gains of approximately $0.3$ in PESQ.
We also analyze the model complexity in terms of the number of trainable parameters.
As shown in Table~\ref{performance_comparison_self}, all proposed methods maintain lightweight model sizes with fewer than $0.3$M parameters.
Among them, the VS-TF NLMS filter requires the smallest number of parameters ($0.067$~M), while the half-recursive and recursive frameworks require $0.168$~M and $0.253$~M parameters, respectively.
Overall, the recursive framework achieves the best overall performance while maintaining a compact model size, whereas the VS-TF NLMS filter significantly improves over conventional adaptive filter-based methods with minimal parameter overhead.

Second, we compare the proposed recursive framework with existing hybrid methods, in terms of ERLE and PESQ, based on the evaluation set.
Specifically, ERLE and PESQ are used to evaluate performance for the single-talk scenarios and the double-talk scenarios, respectively. 
As shown in Table~\ref{performance_comparison}, the proposed recursive framework achieves the best performance among the hybrid methods, yielding highest ERLE and PESQ scores.
Regarding the model complexity, the proposed recursive framework achieves a more lightweight model size with only $0.253$~M parameters, less than that of the NLMSNet method ($0.888$~M parameters) and the KalmanNet method ($0.977$~M parameters).

\begin{figure}[t]
\includegraphics[width=8.5cm]{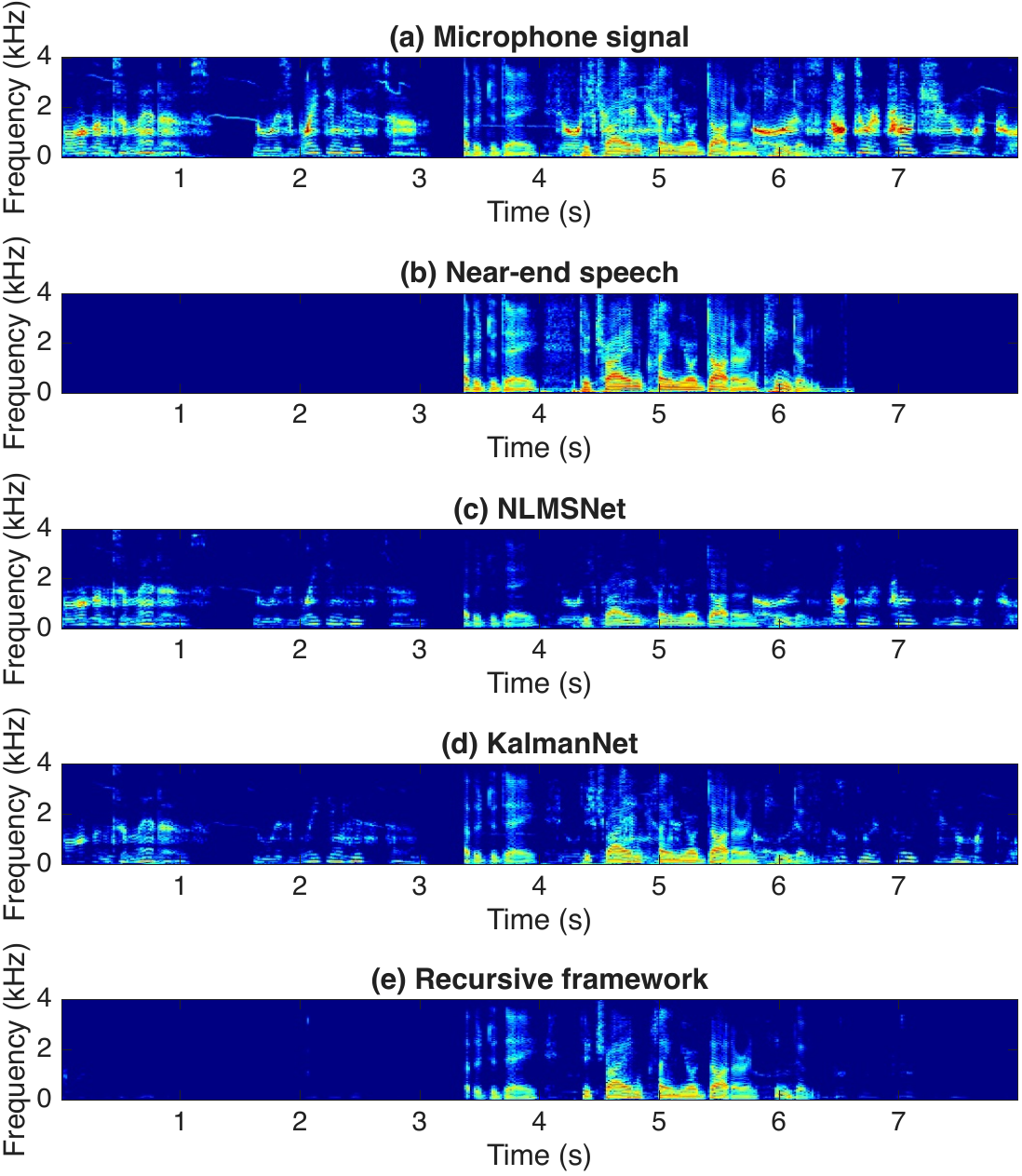}
\centering
\caption{The spectrogram of (a) the microphone signal, (b) the near-end speech, (c) the output of the NLMSNet method, (d) the output of the KalmanNet method, and (e) the output of the proposed recursive framework.}
\label{fig:spectrogram}
\end{figure}

Third, Fig.~\ref{fig:spectrogram} shows the spectrogram of the microphone signal, the near-end speech, and the estimated near-end speech from different methods. 
It is shown that the ability of both the NLMSNet method and the KalmanNet method appears to be limited to suppress echoes.
By contrast, the proposed recursive framework demonstrate its superior effectiveness in removing echoes.
In addition, the NLMSNet method fails to effectively preserve the near-end speech components above $2$~kHz, whereas both the KalmanNet method and the proposed recursive framework achieve better preservation in this frequency range.

\section{Conclusion}
\label{sec:conclusion}
In this paper, we propose a hybrid framework that combines deep learning with a designed VS-TF NLMS filter for AEC and suppression.
A lightweight implementation is adopted with only $253$ k parameters through the proposed constrained recursive training strategy.
In addition, we present a variant as a half-recursive framework with only $168$ k parameters, which first trains part of the neural network parameters recursively and then trains the rest offline. 
Experimental results demonstrate that our proposed frameworks with the proposed VS-TF NLMS filter outperform the baselines.
In the future, we plan to investigate the potential of the proposed frameworks in other audio processing tasks.

\section{Acknowledgment}
The authors would like to thank Jitao Ma and Xing Tian for their work during their internship at Tencent AI Lab.

\bibliographystyle{IEEEbib}
\bibliography{refs}

\end{document}